\documentclass[12pt,preprint]{aastex}

\shorttitle{SSOS}
\shortauthors{Gwyn, Hill and Kavelaars}

\begin{document}


\title{SSOS: A Moving Object Image Search Tool for Asteroid Precovery}


\author{Stephen. D. J. Gwyn, Norman Hill and J.J. Kavelaars}
\affil{
Canadian Astronomical Data Centre,
Herzberg Institute of Astrophysics,
5071 West Saanich Road,
Victoria, British Columbia,
Canada   V9E 2E7
}
\email{Stephen.Gwyn@nrc-cnrc.gc.ca}
\email{Norman.Hill@nrc-cnrc.gc.ca}
\email{JJ.Kavelaars@nrc-cnrc.gc.ca}



\begin{abstract}
It is very difficult to find archival images of solar system objects.
While regular archive searches can find images at a fixed location,
they cannot find images of moving targets. Archival images have become
increasingly useful to galactic and stellar astronomers the last few
years but, until now, solar system researchers have been at a
disadvantage in this respect. The Solar System Object Search (SSOS) at
the Canadian Astronomy Data Centre allows users to search for images
of moving objects. SSOS accepts as input either a list of
observations, an object designation, a set of orbital elements,
or a user-generated ephemeris for an object. It then searches for
images containing that object over a range of dates. The user is then
presented with a list of images containing that object from a variety
of archives. Initially created to search the CFHT MegaCam archive,
SSOS has been extended to other telescope archives including Gemini,
Subaru/SuprimeCam, HST, several ESO instruments
and the SDSS for a total of 6.5 million images.
The SSOS tool is located on the web at: http://www.cadc.hia.nrc.gc.ca/ssos

\end{abstract}


\keywords{
methods: data analysis,
astronomical data bases: miscellaneous,
astrometry,
techniques: photometric
}


\section{INTRODUCTION}
\label{sec:intro}

In many fields of astronomy, image archives are of increasing
importance.  For example, since 2005, more than 50\% of HST papers
have been based on archival data rather than PI
data\footnote{\url{http://archive.stsci.edu/hst/bibliography/pubstat.html}}.
Archival images have become increasingly useful to extra-galactic and
stellar astronomers the last few years but, until now, solar system
researchers have been at a disadvantage in this respect.  While
regular archive searches can find images at a fixed location, they
cannot find images of moving targets.

This is unfortunate, because it could be argued that archival data is
potentially more useful to solar system studies than extra-galactic or
stellar astronomy.  The full scientific potential derived from the
discovery of small solar system bodies (SSSBs) can not be fully
realized until precise orbital parameters for those objects can be
determined. The object must be tracked over successive nights to
obtain an approximate orbit, then over the course of months and years
to refine the object's orbit. But if precovery images containing the
object can be found, and the object unambiguously identified in those
images, the orbit can be rapidly refined by extending the observed arc
into the past.

While a small number of dedicated SSSB surveys have been done and do
include the images needed for tracking such objects, the majority of
ground based imaging surveys do not include a SSSB discovery component.
Indeed, the very requirement of precise orbital determination is
exactly why such surveys do not typically include SSSB discovery.  For
example, over 6000 square degrees of sky have been imaged with the
CFHT MegaPrime camera.  However, only a small fraction of these images
have been carefully searched for SSSBs.  These un-searched images are
likely to provide a rich repository of astrometric measurements for
future SSSB surveys.

The earliest example of the utility of archival data in solar system
studies is the discovery of Neptune. While it is well known that the
search for Neptune was prompted by perturbations in the position of
Uranus, it is less well known that some of the observations of Uranus
that went into these calculations were archival. Uranus had appeared
(misidentified as a star) in various catalogs going back to 1690
before its discovery as a planet in 1781. 
\citet{leverrier1846} and \citep{adams1846} used these critical
observations in their calculations to determine the probable location
of Neptune.

Of course, archival images can be useful even if the object's orbit is
well known. For example, one may want to generate a light curve for an
object. Alternatively, one may want observations of an object at
different wavelengths to measure an object's color.  Or again, in the
case of binary system, the heliocentric orbit maybe well known, but
one may want additional observations to constrain the mutual orbit.

The Solar System Object Search tool (SSOS) at the Canadian Astronomy
Data Centre allows users to search for images of moving objects taken
with a number of telescopes. SSOS accepts as input either a list of
observations, an object designation, a set of orbital elements, or a
user-generated ephemeris for an object. It then searches for images of
that object over a range of dates. The user is then presented with a
list of images containing that object from a variety of archives.

A number of related tools and services already exist.  The ESO
(European Southern Observatory) archive has a
service\footnote{\url{http://archive.eso.org/archive/hst/solarbodies/}}
which allows searches for images of known objects in HST (Hubble Space
Telescope) images. The searches are precomputed. Each HST image is
searched using SkyBoT \citep{skybot} for known objects. However the
tool is limited to the HST and one can not search for new objects.

The Skymorph
service\footnote{\url{http://skyview.gsfc.nasa.gov/cgi-bin/skymorph/mobs.pl}}
allows users to search for HST images, images from the Near Earth
Asteroid Tracking system \citep{NEAT}, Spacewatch \citep{spacewatch1}
images, and a number of older plate catalogs. Input is by object
designation or by orbital elements.

The EURONEAR project \citep[EUROpean Near Earth Asteroids Research]{euronear} provides a service which allows
users to specify Near Earth Asteroids by name and search the Bucharest
Plate archive and the
Canada France Hawai'i Telescope Legacy Survey\footnote{\url{http://euronear.imcce.fr/tiki-index.php?page=Precovery}}
for images.

A tool similar to SSOS has been developed by IPAC, the Infrared
Processing and Analysis Center (Groom, private communication). The
Moving Object Search Tool
(MOST)\footnote{\url{http://irsa.ipac.caltech.edu/applications/wise/$\#$id$=$Hydra\underline{~}wise\underline{~}wise\underline{~}5}}
has been released for use with the Wide-field Infrared Survey Explorer
citep[WISE]{WISE} telescope.  The tool takes as input an object name
or a set of orbital parameters which are converted into an ephemeris of
the object as seen from the spacecraft. MOST then converts that
ephemeris into a series of rectangles on the sky plane which bracket
the ephemeris.  The rectangles are searched using IPAC's highly
efficient positional search. Any images found are further screened
first by time and then by a detailed positional match of the ephemeris
to the image footprints.

The Pan-STARRS project \citep[Panoramic Survey Telescope and Rapid
  Response System]{PanSTARRS} is also developing a similar tool to
search for Pan-STARRS images. (Jedicke, private communication) It has
not yet been released to the public. The tool generates the positions
of all known (or synthetic) objects at the beginning, middle and end
of each night. It then fits a quadratic in (RA, Dec) to the positions
so that to predict the position of the objects as a function of time
through the night.  The tool maintains two kd-trees: one for available
fields and one for predicted positions. When searching for a given
object the field tree is searched and then using the functional motion
fit, it predicts which images the object is likely to lie in. For each
of the likely images, a full n-body ephemeris is done at the time of
the exposure, and the objects position is checked against the image's
coverage.

SSOS has two major advantages over these services: First, SSOS allows
searches by a greater number of input methods, allowing searches for
precovery images of newly discovered objects.  Second, while SSOS was
originally developed for searching for images taken with MegaCam on
CFHT, it has been extended to include images from several other
telescopes.  With the exception of of Skymorph and EURONEAR, other
search tools are currently limited to a single telescope.

Before SSOS can be used, a list of archival images from different
telescopes is compiled, as detailed in section \ref{sec:scrape}.  The
first step of each SSOS query is to convert the user's input into an
ephemeris using one of several methods, as described in section
\ref{sec:input}. Next, SSOS searches along that ephemeris for images
as described in section \ref{sec:match}.

\section{IMAGE HARVESTING}
\label{sec:scrape}

Before any queries can be made, the SSOS image database must be
populated. This is done by going to the various telescope archives and
harvesting the metadata describing each image taken by the telescope.
For each image the SSOS stores the following information:
\begin{itemize}
\item Midpoint of exposure time.
\item RA and Dec of the image center
\item The extent of the image in RA and Dec. Some images are not rectangles and some are not
square to the RA-Dec gridlines. In this case SSOS stores the extrema of the images in both
coordinates. 
\end{itemize}

For some archives ({\it e.g.} HST and most of the MegaCam archive),
the position of the images is known to within an arcsecond or two. For
other archives, the image positions are less well known.  To ensure
that no image is missed in a search due to a faulty position, the
image extent is increased slightly to compensate; a buffer is added in
both RA and Dec. The size of the buffer depends on the typical
astrometric quality of an archive.  A further buffer of 15 arcseconds
is added in both directions to allow for any ephemeris errors.

These buffers increase the completeness but slightly decrease the
purity of the searches.  When searching for a given object, the search
results are unlikely to miss any images containing the object,
but may contain a few false positives. In practise, removing all the
false positives is impossible. For instance, many of the instruments
in the database are mosaic cameras (e.g. MegaCam, SuprimeCam, VISTA),
which have gaps between the detectors. False positives will occur if
an asteroid happens to lie in a gap during the exposure.

These parameters (time, center and extent) are the minimum required to
find the image.  When the World Coordinate System (WCS) of an image is
available and accurate, this is also stored in order to refine the
search as detailed in Section \ref{ssec:resolve},

To speed the searches, a bounding box is generated for each image.
The bounding box completely encloses the image and is described in
integer degrees. Again, to speed the search, the exposure midpoint expressed
as a MJD is truncated and stored as an integer.

 Images which span the celestial meridian have two entries in
the database: one entry describing the part of the of image lying
above RA=0$^\circ$, and the other describing the part of the image
lying below RA=360$^\circ$.  No special provision is made for images
covering the poles, since there are none in the database.

SSOS stores some additional parameters describing each image:
\begin{itemize}
\item Exposure time 
\item Filter 
\item Telescope/Instrument
\item The URL at which the data can retrieved.
\end{itemize}
The first three of the these give some indication of whether the data
will be useful. For example, some users may not want data in a certain
filter, others may only be looking for high resolution data, and short
exposure images might be too shallow for some objects.  The URL will
allow the user to actually retrieve the data once found.

Obtaining the metadata for images stored at the Canadian Astronomy
Data Centre (CADC), where SSOS is based, is relatively easy. The
existing databases describing each image archive are queried directly
and the relevant parameters are ingested into the SSOS database.  In
some cases the headers of the individual FITS images are retrieved to
obtain additional metadata. Currently, the following telescopes and instruments
have been harvested from the CADC:
\begin{itemize}
\item CFHT (Canada-France Hawai'i Telescope): MegaCam, WIRCam and CFH12K
\item Gemini: GMOS images only
\item HST (Hubble Space Telescope): 
WFPC (Wide Field Planetary Camera), 
ACS (Advanced Camera for Surveys) and 
WFC3 (Wide Field Camera 3)
\end{itemize}

Offsite archives must be ``scraped'' over the web. This can take many
forms depending on the archive. The Subaru SuprimeCam image lists are
available as simple ASCII text files.\footnote{{\it e.g.}
  \url{http://smoka.nao.ac.jp/status/obslog/SUP\underline{~}2009.txt}} The
ESO archive can be queried by instrument; while such a query takes
many hours, it will return a list of every image made by that
instrument. Currently the following telescopes and instruments
have been harvested:
\begin{itemize}
\item AAT (Anglo-Australian Telescope): 
WFI (Wide Field Imager)
\item ESO-LaSilla 2.2m: 
WFI (Wide Field Imager) 
\item ESO-NTT (New Technology Telescope): 
EFOSC (ESO Faint Object Spectrograph and Camera), 
EMMI (ESO Multi-Mode Instrument), 
SOFI (SOn oF Isaac), 
SUSI1 and SUSI2 (SUperb Seeing Imagers)
\item ESO-VISTA (Visible and Infrared Survey Telescope for Astronomy): 
VIRCAM (VISTA IR Camera)
\item VLT (Very Large Telscope): 
FORS1 and FORS2 (FOcal Reducer/low dispersion Spectrograph), 
HAWK-I (High Acuity Wide field K-band Imager)
NAOS-CONICA (Nasmyth Adaptive Optics System - COud\'{e} Near-Infrared CAmera
ISAAC (Infrared Spectrometer And Array Camera) and 
VIMOS (VIsible MultiObject Spectrograph) imaging
\item Subaru: SuprimeCam only
\item the Sloan Digital Sky Survey DR8 \citep{sdssdr8}.
\end{itemize}
Images from the NOAO archive, VST (OmegaCam) and the Isaac Newton
Group of telescopes will be added in the future.

In principle, data from any instrument can be ingested into SSOS.  So
far, ingestion efforts have concentrated on image archives from
telescopes with larger apertures or larger fields of view.
Only data taken in broad band filters is included; narrow band data
tends to be too shallow for to be useful for precovery. Finally,
only images taken in optical or near-infrared bands has been
harvested to date. 

\begin{figure}
\plotone{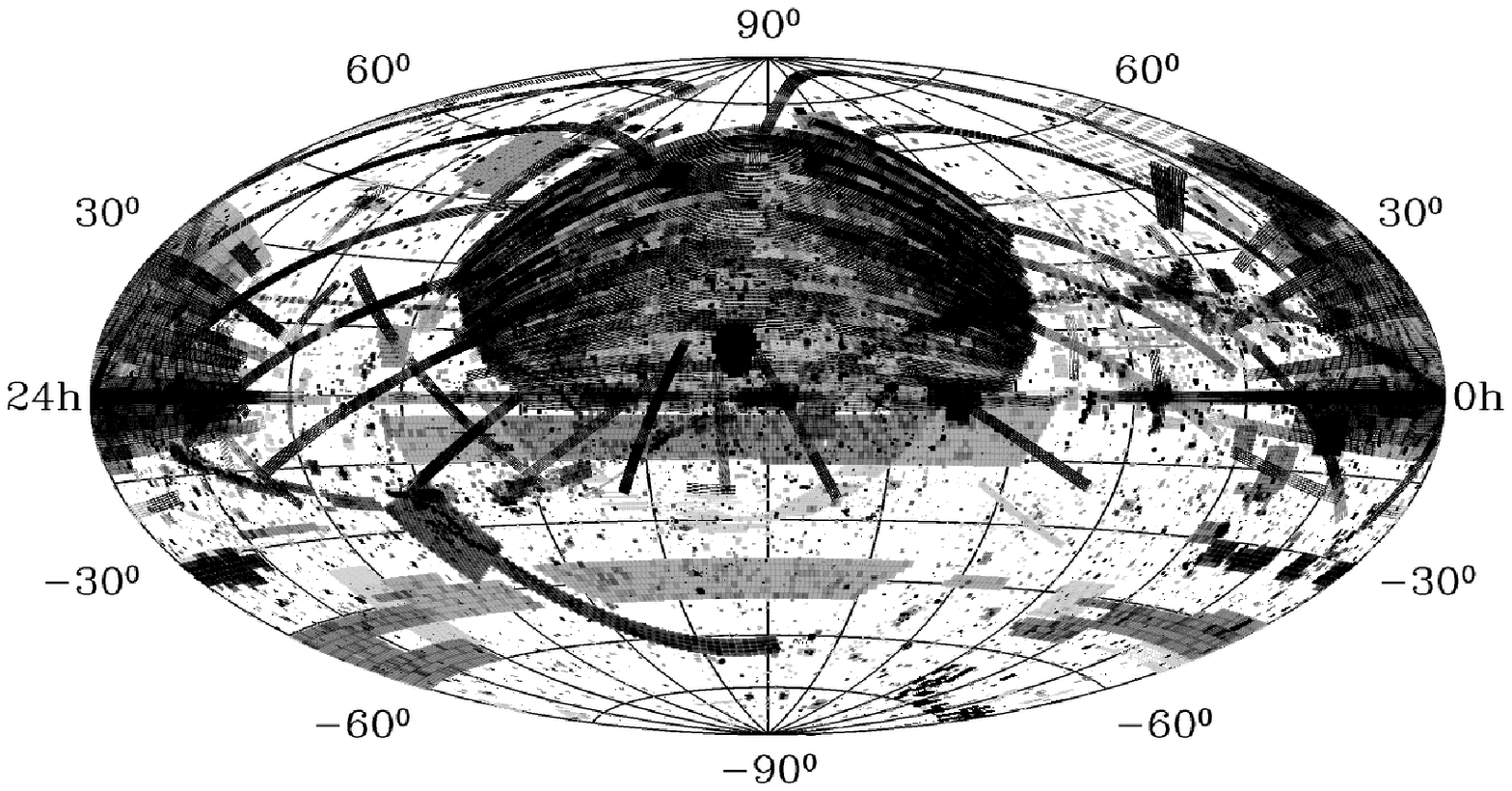}
\caption{Area of the sky covered by SSOS. The greyscale gives an indication the number
of images covering a particular spot on the sky, with a single image being
represented by the faintest grey, and 40 or more images being indicated by solid black}
\label{fig:coverage}
\end{figure}

Currently, there are 1.6 million images in the SSOS database.  Figure
\ref{fig:coverage} shows the area of the sky covered by SSOS.  The
figure indicates the number of images by greyscale. The darkest
patches are covered by at least 40 images. The figure does not
indicate the depth or wavelength of the images. For while a
substantial fraction of the southern hemisphere has been covered
repeatedly by the VIRCAM detector, the exposure time of these infrared
images is typically less than a minute.  As discussed in section
\ref{sec:results}, this coverage is sufficient that two thirds of the
asteroids in the MPC (Minor Planet Center) database lie within at least one image,
and a typical asteroid is covered by 20 images.

\section{USER INPUT AND CONVERSION TO EPHEMERIS}
\label{sec:input}

When arriving at the Solar System Object Search tool
website\footnote{\url{http://www.cadc.hia.nrc.gc.ca/ssos}}, users have four
ways to search for images. In each case, SSOS converts the user's
input into an ephemeris. The four methods of input and conversion are
detailed in the following four subsections:

\subsection{Search by Arc}
In this input method, the user enters a series of observations in MPC
format.\footnote{\url{http://www.minorplanetcenter.net/iau/info/OpticalObs.html}}
SSOS uses these observations to determine an orbit and generate an
ephemeris from that orbit.  The user can select one of two orbit
fitting routines: The orbit fitting code of \cite{bernstein} has been
set up to automatically convert the observations into orbital
parameters (the {\tt fit\underline{~}radec} code) and use those
parameters to produce an ephemeris (the {\tt predict} code). The
Bernstein and Khushalani code works best for outer solar system
objects; if the arc of a main belt asteroid is too short, it tends to
produce spurious results. Therefore, SSOS provides the new object
ephemeris generator from the Minor Planet
Center\footnote{\url{http://minorplanetcenter.net/iau/MPEph/NewObjEphems.html}}
as an alternative.  If a user selects this option, the SSOS queries
the MPC service automatically.  The MPC fits a V\"ais\"al\"a
\citep{vaisala} orbit to the observations and returns an ephemeris
based on this orbit. SSOS then uses that ephemeris.  This method is
slower the the Bernstein \& Khushalani fitting because it requires
SSOS to make a query (often several queries) to an external service.
The ephemeris is generated at intervals of 24 hours.  Mauna Kea
(observatory code 568) is used as the observing site.

\subsection{Search by Object Name}
In this input method, the user enters the name of an object.  SSOS
forwards that name to one of two services, either the Lowell
Observatory asteroid ephemeris
generator\footnote{\url{http://asteroid.lowell.edu/cgi-bin/asteph}} or
the minor planet and comet ephemeris service at the Minor Planet
Center\footnote{\url{http://minorplanetcenter.net/iau/MPEph/MPEph.html}}. These
services query their databases for an object matching the name, make
the appropriate orbital calculations and return an ephemeris to SSOS.
SSOS parses the ephemeris from the external services into a format
which it can use for the image search.

In addition to using these two offsite services, SSOS can also
generate an ephemeris locally. SSOS maintains regularly updated copy
of the MPC orbital element
database\footnote{\url{http://www.minorplanetcenter.net/iau/MPCORB.html}}.
When a user enters an object name, the local version of this database
is queried and the orbital elements are passed to the the program {\tt
orbfit} of \citet{orbfit}\footnote{\url{http://adams.dm.unipi.it/$\sim$orbmaint/orbfit/}}.
which generates an ephemeris.

As with the search by arc option, the ephemeris is
generated at 24 hour intervals and Mauna Kea is used as observing
site.

\subsection{Search by Orbital Elements}
In this case the user enters the orbital elements of an object: epoch,
semi-major axis, eccentricity, inclination, longitude of the ascending
node, argument of perihelion and mean anomaly.  These orbital elements
are used as input to the program {\tt orbfit} which returns an
ephemeris, again at 24 hour intervals and using Mauna Kea as the
observing location.

\subsection{Search by Ephemeris}
\label{ssec:ephem}
This method allows the user complete control over the ephemeris. The
user enters a series of times and object positions. Users can cut
and paste text into the service.  This method is useful if the
user has any concerns about the positional accuracy of any of the
previous methods.  This might be because the object being searched for
is near enough to the earth that the parallax will significantly affect
the objects positions. Alternatively, the object's apparent motion might
be irregular enough that the the linear daily interpolation scheme is
insufficiently accurate. Finally, the user might have concerns about
the accuracy of the program generating the ephemeris.

\subsection{Ephemeris Accuracy}
\label{ssec:ephac}
For the first three types of searches, the ephemeris is generated with
Mauna Kea as the reference location because two of the most useful
instruments in the SSOS database are located there. In this sort of
search, usefulness can be quantified as mirror area $\times$
field-of-view $\times$ years in operation. MegaCam on CFHT and
SuprimeCam on Subaru are the clear leaders.  Mauna Kea is also the
location of Gemini North. However the rest of the telescopes are at
other locations on the earth, while HST is in orbit. How much does the
assumption of Mauna Kea as the observing location affect the
ephemeris?

For main belt asteroids, the assumption will introduce errors of at
most $\sim$10 arcseconds. The parallax error is on the order of 1
earth radius divided by 1 AU (approximately the closest approach of a
main belt asteroid), which is 8 arcseconds. For more distant objects,
such as Kuiper Belt Objects, the error will be significantly less. In
the extreme case of near-earth objects, the parallax error can be much
greater, but for short periods. In the vast majority of cases, the 15
arcsecond buffer described in Section \ref{sec:scrape} is sufficient.

Again, for the first three types of searches, the ephemeris is
generated at 24 hour intervals. For times between these points, linear
interpolation is used to determine the location of the object, as
detailed in the following section.  This is done for reasons of
speed. But how much is the ephemeris accuracy affected relative to ---
for example --- an hourly ephemeris? 

This was tested by generating ephemerides at 15 minutes intervals for
a few key objects: a KBO, a main belt asteroid and an Apollo: Pluto
(134340), Ceres (1) and Eger (3103). The positions were compared to
the results of the linear interpolation described in Section
\ref{sec:match}. The size of the resulting offsets were smaller than
the parallax errors above: less than an arcsecond for for KBOs, a few
arcseconds for main belt asteroids and, while generally small for the
Apollos, occasionally as large as 2-3 arcminutes.  Again, the buffers
added to the image extents described in Section \ref{sec:scrape}
should be sufficient in most cases.

In exceptional cases, users will have to use the direct ephemeris
input method described in section \ref{ssec:ephem}.  Alternatively, if
an increased false positive rate is acceptable, they can increase the
positional uncertainty as described in the following section.

\section{ADDITIONAL INPUTS}
\label{sec:add}

\subsection{Positional Uncertainty}
\label{ssec:poserr}
In addition to ephemeris information, the user can also specify
positional uncertainty information. This is useful if an object's
ephemeris is not perfectly known.  Instead of searching only for
images containing the object's assumed location, the search is
broadened to include images nearby. The search is increased from a
point to a search box whose size is specified by the user.  In most
cases, the size of the search box is fixed.  However, if an arc search
using the Bernstein \& Khushalani code is selected, then the user can
choose to use the error ellipse generated by the code. The search box
at each location grows to contain the error ellipse.

\subsection{Positional Resolution}
\label{ssec:resolve}
The MegaPrime mosaic camera on CFHT consists of 36 separate CCDs.
Each MegaCam image is large: 700Mb (300Mb in compressed
format). Downloading a large number of the these images can be time
consuming depending on the user's bandwidth. However, the images are
stored in Multi-Extension FITS format.  It is possible to download a
single extension (corresponding to a separate chip) from the CADC. At
about 16Mb in size, it offers an attractive alternative to the full
image.  Therefore, SSOS offers an option to resolve the search down
the extension level for MegaCam images. This option is only available
for the MegaCam images which have been calibrated using the MegaPipe
pipeline, currently about 75\% of the total MegaCam archive. When this
option is active, the search returns a link to the image extensions
which likely contain the object being searched for. If the positional
uncertainty is specified (as in Section \ref{ssec:poserr} above) links
to all the CCDs spanned by the error box are returned.

Similarly, SSOS also has the option to resolve down to the pixel
coordinates within a single chip. The stored world coordinate system
for that chip is used to convert the RA and Dec of the object to an XY
pixel location which is returned the user. 

\section{SEARCHING ALONG THE EPHEMERIS}
\label{sec:match}

Once the first step, that of generating an ephemeris, has been
accomplished, the next step is to match that ephemeris to the database
of images.  The ephemeris is converted to a temporary database
table. Each interval in the ephemeris becomes a row in the table, with
start and end times and positions. A bounding box is generated for
each row, covering the full span in time and position. For speed, this
bounding box is expressed as integer days (for the time) and integer
degrees (for position).  When the object moves across the first point
of Aries, two row are added, one each describing the position of the
object on either side of the celestial meridian.  If the time interval
spans multiple days (for example if the user generated ephemeris has
been produced at weekly intervals), additional rows, one per day, are
added to this table.  This temporary table is comparatively small. If
the ephemeris is sampled daily, just under 8000 rows suffice to cover
the time span from the earliest image in the SSOS database to
the present. Building the temporary table takes 2-3 seconds for a full
20 year span.

The ephemeris table is then cross matched to the image table; in
the terminology of relational databases, they are ``joined''.  The
integer bounding boxes of the ephemeris and the images are
matched first. If the bounding boxes of an ephemeris and an
image match, the object's position is calculated more accurately
at the image's exposure midpoint by linearly interpolating the
ephemeris. The linear interpolation is key to keeping the queries
reasonably fast. Doing a full orbital prediction for each of the
images is not feasible. This is sufficiently accurate for the majority
of queries, where the object either moves slowly or in a fairly
straight line. For faster moving, nearby objects, it may be necessary
to supply the ephemeris sampled at shorter time steps, as discussed in
section \ref{ssec:ephac}.  If a spline method is used (rather than
linear interpolation), the positional accuracy increases, but only by
a few percent and at considerable computational cost.  A typical cross
match takes 0.3 seconds to match a 20 year ephemeris to a 1.6 million
images.

\section{RESULTS AND PERFORMANCE}
\label{sec:results}
The results page presents the user with a table listing the image name,
exposure midpoint, filter name, exposure time, the objects position at
exposure midpoint, the image target name and links to more metadata.
If possible (for example if the image is hosted at the CADC), SSOS
provides a direct HTML link to the image. For data centers such as
SMOKA and ESO, which operate on the request-stage-retrieve model, SSOS
provides a link to the request webpage for that image.

In addition to providing a table, SSOS provides a plot showing the
location of the object on the sky, with a line showing the objects
location and dots showing matching images. This information is also
made available as a ds9 regions file.

The speed of the search depends on which ephemeris method is used and
the time span being searched.  The fastest example is if the orbit
fitting method of Bernstein and Khushalani method is used and the time
span is short: a few months before and after the input observations.
This would the span useful for precovery of a newly discovered object.
In this case, the search would require less than a second to generate
the ephemeris, build the temporary table and match to the SSOS
image table. On the other hand, if one searches by object name
sets the span to the full twenty year span of the SSOS image database,
and requests that the MPC be used to generate the ephemeris, the
search will take longer. The MPC queries can take 10 or more seconds
(depending on the load on the MPC servers), it will take 2-3 seconds
to generate the temporary ephemeris table, and the cross matching will
take a second. The represents the slowest possible search.

\begin{figure}
\plotone{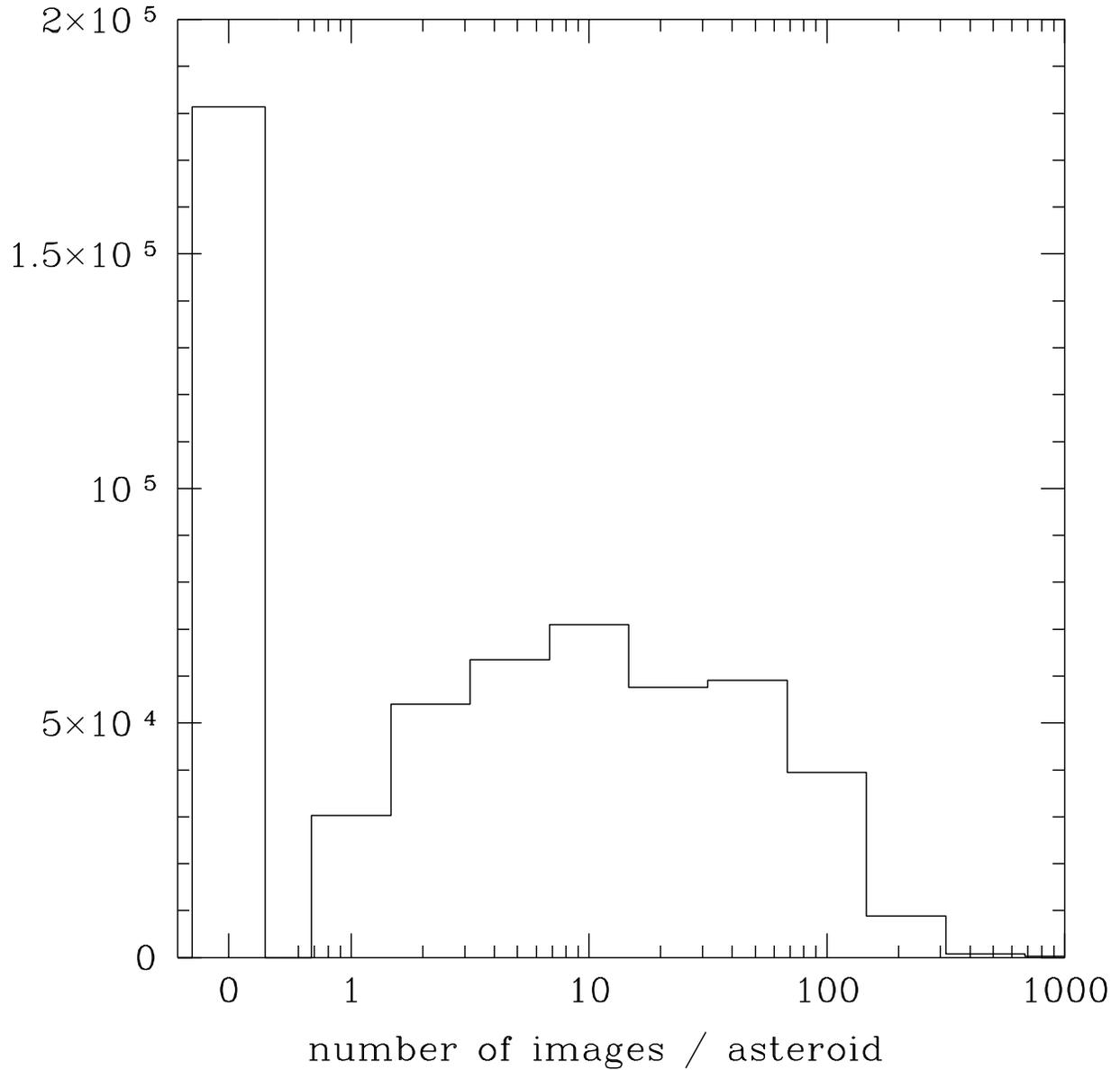}
\caption{Number of images containing an asteroid. The figure was generated by
feeding all the objects in the MPC orbital element database into SSOS
and plotting the number of returned images as a histogram. Note that
the axis is logarithmic. Searches returning no images are shown as a
separate histogram bar. Two thirds of the searches return at least one
image.  On average. an image search for given object will return 10
images, but with considerable scatter. }
\label{fig:orbmatch}
\end{figure}

How likely is it that SSOS will find images of a particular object?
And how many images of an object is it likely to find?  Obviously, if
a typical search does not return any images, the service is not very
useful. This was investigated by searching for images of 566~253
objects from the MPC orbital element database. It was found that, of
these searches, 384~786 returned at least one image, {\it i.e.} just
over over two thirds of the searches were successful.  The results are
shown in figure \ref{fig:orbmatch}. The number of images returned by
the searches is shown as a histogram. Searches returning no images are
shown as a separate histogram bar.  In some cases, several hundred
images are returned. If only the successful searches ({\it i.e.} the
searches returning at least one image) are considered, the average
number of image returned is 30. The average number of images returned
over all searches is 20 (including the third of the searches that
returned nothing).

Note that this is just the number of images returned, which may not be
a good indication of how many useful observations may be derived from
these images.  SSOS will only find the images containing an object.
It does not in general find the position of that object in the images.
Where the orbital parameters are poorly known, positional uncertainty
may make it difficult to find the object in question.  Indeed, if the
position is poorly known and the image's field of view is small, all
that SSOS can say is that the image probably contains the object. If
there are many moving objects in the images, it may be hard to
unambiguously identify the object in question. Unless there are
multiple images of the object close together in time, it will be hard
or impossible to identify the object. Alternatively, there may be
enough images coincident in position and time, but taken in different
wavelengths, making it harder to identify the moving object. As noted
in section \ref{sec:scrape}, some of the instruments described in SSOS
are mosaic cameras with gaps between the detectors; SSOS does not
describe the footprint of the cameras in detail, so objects may fall
between the gaps. Finally, there is no guarantee that the images
returned are deep enough to detect the object, even if it covers the
correct patch of sky.

\section{USAGE}

To date, SSOS has been used successfully in a number of projects:

\citet{tnobinary} used the tool to locate additional data of wide
binary TNOs. SSOS found images in the HST and Gemini archives that
were used to constrain the mutual orbit of the binaries.

The SSOS tool was used by the Canada-France Ecliptic Plane Survey
\citep[CFEPS]{cfeps} project in two ways.  First, they used the tool
to retrieve the survey's images of particular KBOs.  This allowed the
rapid re-analysis of the discovery data to search for binaries among
the CFEPS KBOs \citep{2010PASP..122.1030L}.  Secondly, the ability to
easily return to the search data allows members external to the
original survey to enhance the survey dataset.  The search tool was
also used to look for, and find, pre- and post- discovery images of
the KBOs discovered by the CFEPS project.

SSOS is also part of the KBO search in the Next Generation Virgo
Survey \citep[NGVS]{ngvs1}. While the main purpose of the NGVS is the
study of the properties of galaxies in the Virgo cluster, the same
data is being used to search for KBOs. SSOS is used to search for
additional images of the newly detected KBOs.

Alexandersen et al (in preparation) discovered a Jovian moon in
2010. Using SSOS and its user-generated ephemeris feature, they were
able to find precovery images dating back to 2003 from the CFHT
MegaCam archive.

Finally, Fraser et al. (in preparation) used SSOS for precovery of
Kuiper belt objects as part of a follow-up program of objects
discovered in Fraser, Brown, and Schwamb
\citep{2010Icar..210..944F}. These targets are part of a large sample
used to determine the size distribution of different Kuiper belt
dynamical populations, which will help constrain the formation history
of the outer Solar system.

\clearpage

\clearpage

\end{document}